\documentclass[prd,reprint,nofootinbib]{revtex4-1}
\usepackage{fullpage}

\usepackage{xparse}
\usepackage{color}
\usepackage{amsmath}
\usepackage{amsfonts,amssymb}
\usepackage{slashed}            
\usepackage{bbm}                
\usepackage{xspace}				
\usepackage{mathrsfs}
\usepackage{hyperref}
\usepackage{cprotect}
\usepackage{subcaption}

\numberwithin{equation}{section} 
\usepackage{epstopdf}

\usepackage{tikz}
\usetikzlibrary{arrows,shapes}
\usetikzlibrary{trees}
\usetikzlibrary{matrix,arrows} 				
\usetikzlibrary{positioning}				
\usetikzlibrary{calc,through}				
\usetikzlibrary{decorations.pathreplacing}  
\usetikzlibrary{decorations.pathmorphing}	
\usetikzlibrary{decorations.markings}

\usepackage{fancyvrb}

\newcommand{\fvc}{F\!\!\!\!$/$\,-C\,\,}
\newcommand{\fvs}{F\!\!\!\!$/$\,-S\,\,}
\newcommand{\fv}{F\!\!\!\!$/$\,\,}
\newcommand{\fc}{F$_\text{C}$\,\,}
\newcommand{\fcc}{F$_\text{C}$-C\,\,}

\newcommand{\mtone}{m_{\tilde{t}_1}}
\newcommand{\mttwo}{m_{\tilde{t}_2}}
\newcommand{\mbone}{m_{\tilde{b}_1}}
\newcommand{\mchi}{m_{\tilde{\chi}^0}}
\newcommand{\thetat}{{\theta_{\tilde{t}}}}

\newcommand{\tone}{\tilde{t}_1}
\newcommand{\bone}{\tilde{b}_1}
\newcommand{\ttwo}{\tilde{t}_2}
\newcommand{\neut}{\tilde{\chi}^0}
\newcommand{\ETmiss}{E_T^\text{miss}}

\usepackage{braket,bm}
\usepackage{subcaption,mathtools,feyn}
\usepackage{hhline} 
\usepackage{cancel}

\newcounter{qnumber}

\newcommand{\Fig}[1]{Fig.~(\ref{#1})}
\newcommand{\Sec}[1]{section~\ref{#1}}
\newcommand{\Equ}[1]{eq.~(\ref{#1})}
\newcommand{\Tab}[1]{table~(\ref{#1})}

\begin{document}

\title{Mixed Stops and the ATLAS on-$Z$ Excess}
\author{Jack H Collins}
\email{jhc296@cornell.edu}
\author{Jeff Asaf Dror}
\email{ajd268@cornell.edu}
\author{Marco Farina}
\email{mf627@cornell.edu}
\affiliation{Department of Physics, LEPP, Cornell University, Ithaca NY 14853}
\begin{abstract}
The ATLAS experiment has recently observed a $3\sigma$ excess in a channel with a leptonically decaying $Z$, jets, and $E_T^\text{miss}$.  It is tantalizing to interpret the signal as the first sign of a natural supersymmetric spectrum. We study such a possibility in a minimal model containing light stops and a neutralino LSP. The signal is characterized by a novel topology (compared to previous attempts) where the $Z$ is emitted from a colored particle in the first step of a decay chain, namely $\tilde{t}_2 \to \tilde{t}_1 Z$ which is characteristic of mixed stops. We show that the excess is compatible with a compressed stop spectrum and is not excluded by any other relevant search, finding some regions of parameter space with signal strength within $1\sigma$ of that measured by the ATLAS collaboration. In addition, we notice that the corresponding CMS search could be prone to background contamination in unexpected topologies of this kind.
\end{abstract}

\maketitle
\section{Introduction}
Supersymmetry (SUSY) is a leading candidate for resolving the large hierarchy problem of the Standard Model. In the Minimal Supersymmetric Standard Model (MSSM) and simple extensions, a necessary feature for a complete resolution of the hierarchy problem is the presence of two light (sub-TeV) colored stops and one light left handed sbottom (to accompany the left handed stop). A common assumption in these models is an exact R-Parity, and the presence of a neutral, stable lightest Supersymmetric particle (LSP). In this case, if the third generation squarks are accompanied by a neutralino LSP, $\chi^0$, then typical decays of these particles include $\tilde{t}_{1,2} \to t^{(*)} \neut $ (where the superscript on $t$ indicates the possibility that it is off-shell), $\tilde{t}_2 \to \tilde{t}_1 Z$ and $\tilde{b}_1 \to b \neut$. The signatures of this scenario are therefore jets, missing transverse momentum ($\ETmiss$), leptons and $b$-tagged jets.

Dedicated searches for 3rd generation squarks have found no deviations from SM predictions, placing stringent constraints on its parameter space. On the other hand there remain significant windows allowing the mass of the lightest stop $\mtone$ to be as light as 200 GeV, provided that there is a compressed spectrum which softens the $p_T$ distributions of the final state particles. Intriguingly, a recent ATLAS search found a $3 \sigma$ excess in final states containing a leptonically decaying $Z$ boson, jets, and large $E_T^\text{miss}$~\cite{ATLAS_Z}. They found 29 events in a combined signal region with expected SM background of $10.6 \pm 3.2$ events. We wish to explore the possibility that this excess is a first signal for direct production of $t_2$ followed by the decay $t_2 \to t_1 Z$\footnote{See~\cite{Ghosh:2013qga} for other recent work on this signature}.

Various attempts have been made to explain this excess in terms of SUSY models\footnote{See~\cite{Vignaroli:2015ama} for a discussion of this excess in the framework of Composite Higgs models.}~\cite{Ellwanger:2015hva, Cao:2015ara,Allanach:2015xga,Barenboim:2015afa,Kobakhidze:2015dra,Cahill-Rowley:2015cha,Lu:2015wwa,Liew:2015hsa,Cao:2015zya}. In all of these studies, pair produced colored particles (squarks or gluinos) decay into quarks and an uncolored particle, which then decays into a $Z$ boson and an LSP. The principal challenge faced by these models is in explaining the ATLAS excess while simultaneously evading the many bounds imposed by other searches by ATLAS and CMS for multileptons or jets and $\ETmiss$, as well as a similar CMS search for the same final state~\cite{CMS_Z} which saw no excess over Standard Model (SM) backgrounds. This latter search imposed different cuts from the ATLAS search and so does not necessarily rule out new physics explanations for the ATLAS excess, yet it still imposes stringent constraints (see, e.g. \cite{Allanach:2015xga}).

The phenomenology of the signal proposed in this paper differs from that of the aforementioned possibilities in several key respects. Firstly, the topology differs in that the $Z$ boson is emitted at the first stage in the decay, rather than at the end with the LSP. This opens up the possibility that the CMS search is subject to significant background contamination, as we discuss at the end of \Sec{sec:scan}. More significantly, our scenario requires the presence of three new colored particles in the spectrum which are lighter than in previous explanations, the heaviest of which gives rise to the desired signature. 
Evading dedicated searches for these particles places very particular constraints on the mass splittings and decays of the squarks. As we shall discuss in \Sec{sec:Model}, this requires a compressed splitting between $\tilde{t}_1$ and $\neut$, and possibly also between $\tilde{t}_1$ and $\tilde{t}_2$. This in turn motivates the consideration of flavor violating decays of $\tilde{t}_1$ into $u \neut$ or $c \neut$, resulting from mixing between the right handed squarks. Such mixings have been discussed in recent years motivated by the question of natural SUSY and light stops~\cite{Blanke:2013uia, Backovic:2015rwa, Agrawal:2013kha}, but without the $Z$ decay necessary to explain this excess.

In \Sec{sec:Model} we provide a systematic discussion of the possibilities that this minimal stop scenario affords for explaining the excess, identifying three distinct scenarios characterized by the mass splittings involved and the assumed decay mode for the light stop. We proceed in \Sec{sec:searches} to describe the main experimental searches placing limits on these scenarios, and perform scans of their parameter spaces to find regions in which the excess can be explained while evading those limits in \Sec{sec:scan}. We also note the possibility that the signal topologies that we have identified could cause significant background contamination in the CMS on-$Z$ search.

 \begin{figure*}[t]
 \centering
 \begin{subfigure}{0.49\linewidth}
 \hspace{-10 mm}
 \includegraphics[width=0.8\linewidth]{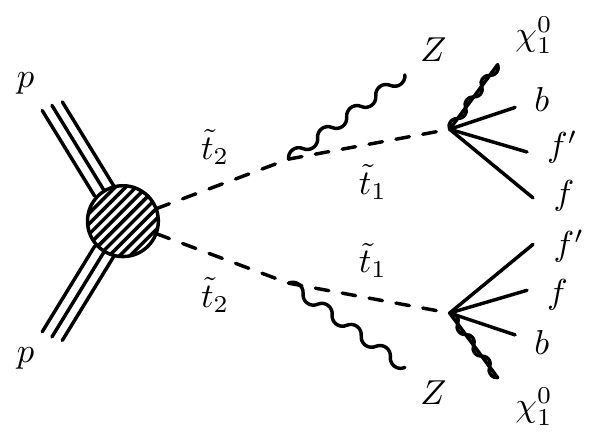}
 \end{subfigure} \hfill
  \begin{subfigure}{0.49\linewidth}
  \hspace{-10 mm}
  \includegraphics[width=0.8\linewidth]{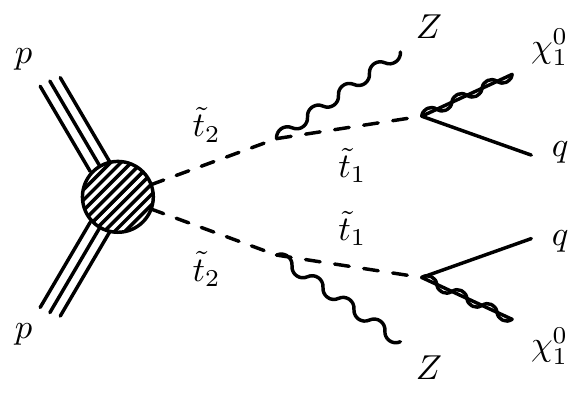}
 \end{subfigure}
 \caption{Flavor conserving (left) and flavor violating (right) decays contributing to the on-$Z$ excess. In the flavor violating case, $q$ can be either an up or a charm quark.}
 \end{figure*}

\section{Model Overview}\label{sec:Model}

In this paper we assume a minimal model including two light stops, $\tone$ and $\ttwo$, and a left-handed sbottom $\bone$. We also require one neutralino LSP, $\neut$. As will be discussed below, the identity of this neutralino is not relevant to collider phenomenology for two of the scenarios that we consider, while for the third case it will be assumed to be mostly Bino\footnote{We do not want to address any cosmological issue in this work but let us notice that a stable Bino is overproduced in the early universe according to the usual thermal freeze-out calculation. On the other hand, a small $\tone \text{--} \neut$ mass splitting of $\mathcal{O}(30 \; \text{GeV}$ allows for the possibility that the correct relic density results from stop-neutralino coannihilation \cite{Boehm:1999bj,Balazs:2004bu}. Alternative solutions simple solution are to assume a low reheating temperature or that that the Bino is actually the NLSP (for instance with a gravitino LSP) but still stable on detector lengths.}. All other SUSY states are assumed to be heavier than these particles and decoupled.  In the MSSM, the stop and sbottom mass matrices in the gauge-eigenbasis are given by~\cite{Martin}:
\begin{align}
\mathbf{m^2_{\tilde{t}}} = \begin{pmatrix}m_{Q_3}^2 + m_t^2 + \Delta_{\tilde{u}_L} & v \left( a_t^* s_\beta - \mu y_t c_\beta \right)\\  v \left( a_t s_\beta - \mu^* y_t c_\beta \right) & m_{u_3}^2 + m_t^2 + \Delta_{\tilde{u}_R} \end{pmatrix},\\
\mathbf{m^2_{\tilde{b}}} = \begin{pmatrix}m_{Q_3}^2 + \Delta_{\tilde{d}_L} & v \left( a_b^* c_\beta - \mu y_b s_\beta \right)\\  v \left( a_b c_\beta - \mu^* y_b s_\beta \right) & m_{d_3}^2  + \Delta_{\tilde{d}_R} \end{pmatrix},
\end{align}
where $m _{ Q _3 }$, $m _{ u _3 }$, $m _{ d _3 }$, $a _t$ , $\mu$ are soft SUSY breaking parameters, $c_\beta$ and $s_\beta$ denote the cosine and sine of $\beta$, and  $\Delta_{\tilde{q}} = (T_{3 {\tilde{q}}} - Q_{\tilde{q}} \sin^2 \theta_W)c_{2 \beta}  m_Z^2$ with $ T  _{3 \tilde{q}}$ and $Q _{ \tilde{q} } $ denoting the third component of weak isospin and electric charge respectively. The Higgs vev $v$ is $ \approx  174  \mbox{GeV} $. We assume the right handed sbottom is decoupled, with $m_{\tilde{d}_3}^2 \gg m_{Q_3}^2$. We replace the MSSM parameters in these mass matrices with physical parameters: the stop mass eigenstates $m_{\tilde{t}_1}$, $m_{\tilde{t}_2}$, and the mixing angle $0 < \thetat < \pi/2$ which rotates the gauge-eigenstate basis into the mass basis
\begin{equation}
\begin{pmatrix} \tilde{t}_1\\ \tilde{t}_2 \end{pmatrix} = 
\begin{pmatrix}
c_{\thetat} & -s^*_{\thetat}\\
s_{\thetat} & c_{\thetat}
\end{pmatrix}
\begin{pmatrix} \tilde{t}_L\\ \tilde{t}_R \end{pmatrix}.
\end{equation}
The sbottom mass is then given by
\begin{equation}
m_{\tilde{b}_1}^2 = \mtone^2 c_{\thetat}^2 + \mttwo^2 s_{\thetat}^2 - m_t^2 - \Delta_{\tilde{u}_L} + \Delta_{\tilde{d}_L}.
\label{eq:mb1}
\end{equation}

We assume the decoupling limit for the Higgs sector, so that the Higgs mixing angle $\alpha = \beta - \pi/2$. The phenomenology of this simplified model varies only slightly with $\tan \beta$ and we therefore choose to fix $\tan \beta = 20$. The remaining free parameters in the model are $\mtone$, $\mttwo$, $\mchi$, and $c_\thetat$. Even with such a modest amount of new particles, this model admits a rich phenomenology with many possible final states, depending mainly on the assumed mass splittings and mixings involved. We seek scenarios with a large branching ratio (BR) for $\ttwo \to \tone Z$, and which are poorly constrained by dedicated searches for $\tone$ and $\bone$. In the following subsections, we systematically discuss the various possibilities and present a categorization of interesting scenarios based on the assumptions made about the decays of $\tone$ and the mass splitting between $\tone$ and $\ttwo$.

\subsection{$\mathbf{\tone}$ decays}

The strongest constraints on the $\tone$ apply if it decays directly to a neutralino and on-shell top, leading to final states with large $\ETmiss$, hard $b$-jets, and leptons. We therefore take the splitting $\mtone - \mchi < m_t$, such that the only flavor-conserving decays that are kinematically available to the light stop are into the three- or four-body final states $W b \neut$ or $f f' b \neut$ (where $f f'$ are pairs of fermions that may be produced in the decay of an off-shell $W$). This allows $\tone$ to be as light as 300~GeV for generic values of this splitting, and as low as 200~GeV in some narrow windows of parameter space (see~\cite{Aad:2015pfx} for a detailed discussion).
 
Due to the substantial kinematic suppression of the partial width into these states, it is possible that flavor violating decays might dominate even with small couplings. This motivates our consideration of flavor violating decays. One well explored possibility arises even with Minimal Flavor Violation (MFV)~\cite{D'Ambrosio:2002ex}, in which case it is possible that loop-induced decays into charm and neutralino can dominate over four-body decays~\cite{Muhlleitner:2011ww}. In recent years an alternative scenario has been explored, that non-MFV mixings between right handed up-type squarks can substantially alter stop phenomenology. Briefly, the essential point for our analysis is that the strongest constraints on the size of squark flavor mixings from low energy observables apply to the down sector, and on mixings between up and charm squarks. The constraints on the down sector also impose constraints on the left-handed up type squarks. Crucially, there are no direct constraints on $\tilde{t}_R-\tilde{c}_R$ or $\tilde{t}_R-\tilde{u}_R$ mixings individually, but only on their product (coming from the $D^0 -\bar{D}^0$ system). We refer readers to the papers~\cite{Blanke:2013uia,Backovic:2015rwa, Agrawal:2013kha} for a more detailed discussion. As a consequence, there may be size-able mixing between $\tilde{t}_R$ and $\tilde{c}_R$ or $\tilde{t}_R$ and $\tilde{u}_R$, but not both.

The degree of flavor mixing can be parameterized by the quantity $\epsilon \equiv (m^2_{\tilde{u}})_{i 3}/(m^2_{\tilde{u}})_{i i}$, where $m^2_{\tilde{u}}$ is the up-type squark mass matrix in the Super-CKM basis and $i$ is 1 or 2. We do not require $\mathcal{O}(1)$ mixings in order to change the decay patterns of the lightest stop, so long as it has an $\mathcal{O}(1)$ admixture of $\tilde{t}_R$. In particular, in the four-body region\footnote{Even when talking about flavor violating two-body decays, we find it convenient to label the regions of parameter space in the $\tone, \neut$ plane by the possible flavor conserving decays. The `three-body region' is defined by $m_W + m_b < \mtone - \mchi < m_t$, while the `four-body region' is defined by $m_b < \mtone - \mchi < m_W + m_b$.}, $\epsilon \gtrsim 10^{-3}$ is sufficient for the decay $\tone \to q \neut$ to occur at least 90\% of the time, while $\epsilon \gtrsim 10^{-2}$ is sufficient for much of the three-body region. These flavor-mixing angles are sufficiently small not to play a noticeable role in the phenomenology of the $ \tilde{t} _2 $, and are relevant for $ \tilde{t} _1 $ only because its flavor-conserving decays are heavily suppressed. It can also be assumed that $(m^2_{\tilde{u}})_{i i} \gg (m^2_{\tilde{u}})_{3 3}$, such that despite introducing a small admixture of first or second generation squark flavor into the two dominantly stop mass eigenstates, the other mass eigenstates are beyond reach of the first run of the LHC.

We therefore consider separately scenarios where $\tone$ undergoes a flavor-conserving decay (denoted \fc), or a flavor violating one (\fv). In either case we will assume a 100\% BR for the light stop for simplicity, though as pointed out in~\cite{Grober:2014aha} the limits on light stops can be substantially reduced for mixed \fv and \fc decays. The flavor violating decay may be into $u \neut$ or $c \neut$, but not both. The only difference this makes regarding collider phenomenology is that the constraints on final states containing charm quarks are often stronger than on up quarks, due to significant progress made on charm flavor tagging by the experimental collaborations. This will be discussed in more detail in \Sec{sec:searches}. It should be noted that the precise measurement of the neutron Electric Dipole Moment (EDM) places constraints on $\tilde{t}_R -\tilde{u}_R$ mixing in the presence of large stop L-R mixing as exists in our model. This comes from loop contributions to the up quark EDM involving gluinos, and depends sensitively on details about the particles which we have assumed to be decoupled in our scenario. Nonetheless, it is demonstrated in~\cite{Dedes:2015twa} that $\epsilon \lesssim 10^{-2}$ can be consistent with the EDM constraints without making unnatural assumptions about the masses of the other particles, or about cancellations between different contributions.

\subsection{$\mathbf{\ttwo}$ decays}
\begin{figure}
\centering
\includegraphics[width=0.9\columnwidth]{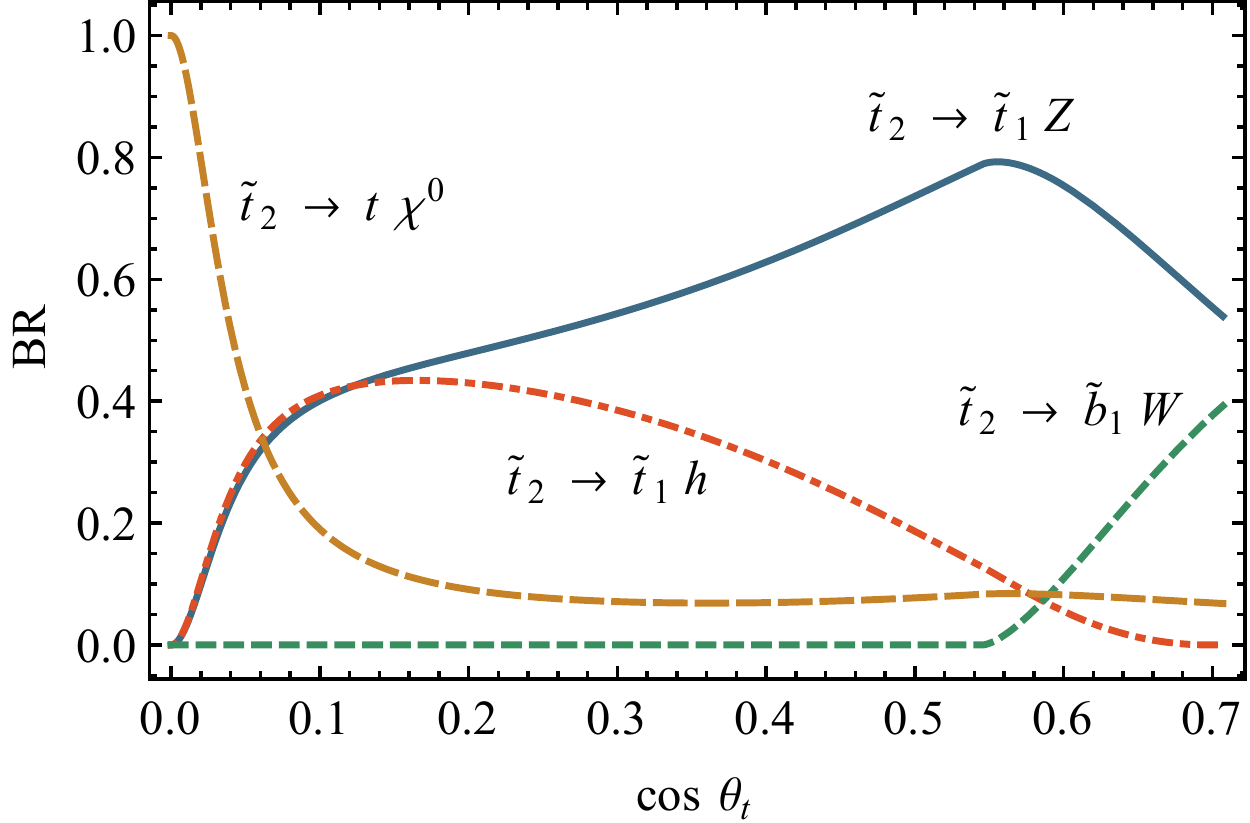} 
\caption{Heavy stop branching ratios in the split scenario, for $\mttwo = 450 \; \text{GeV}$, $\mtone = 250 \; \text{GeV}$, $\mchi = 210 \; \text{GeV}$, $\tan \beta = 20$.}
\label{fig:BRs}
\end{figure}

The second important distinction to be made between different classes of scenarios relates to the mass splitting between $\ttwo$ and $\tone$, and the role that this plays in determining the branching ratios of the three squarks. For sufficiently large mass splittings, the possible two-body decays of the heavy stop are $\ttwo \to \tone Z$, $\ttwo \to \tone h$, $\ttwo \to \bone W^+$ and $\ttwo \to t \neut$. The BRs into these states are most sensitive to the mixing angle $c_\thetat$, and for splitting $150 \; \text{GeV} \lesssim \mttwo - \mtone \lesssim 300 \; \text{GeV}$ the BR into $Z \tone$ is maximized at a value between $0.6 \lesssim \text{BR}(\ttwo \to \tone Z) \lesssim 0.8$ for $0.5 \lesssim c_\thetat \lesssim 0.55$, as illustrated in \Fig{fig:BRs}. Since we are interested in maximizing the $ \tilde{t} _2 \rightarrow \tilde{t} _1 Z $ BR we can use this to fix the mixing angle. This class of scenario, which we label `split', has a three-dimensional parameter space in $\mtone$, $\mttwo$ and $\mchi$. Note that for small $ \cos \theta _t $, $ \tilde{t} _1 $ and $ \tilde{b} _1 $ are almost degenerate and hence the decay $ \tilde{t} _1 \rightarrow \tilde{b} _1 W $ is not kinematically available. 

An interesting alternative is that the $\ttwo-\tone$ mass splitting is sufficiently small that the only two-body decay kinematically allowed for $\ttwo$ is $\ttwo \to \tone Z$. For $\mtone - \mttwo < m_h$, the decay $\ttwo \to \tone h$ is forbidden. For a wide range of $c_\thetat$, the decay $\ttwo \to \bone W^+$ is not kinematically available. Finally, if $m _{ \tilde{t} _1 } + m_Z < m _{ \tilde{t} _2 }< m_t + \mchi$, the only two-body decay that is kinematically available is the $Z$ decay. Combining the bounds gives the requirement that $ m _{ \tilde{t} _1 } - \mchi < m _t - m _Z \approx   85 \; \text{GeV}$. Coincidentally, this turns out to overlap almost exclusively with the four-body region, which is defined by $ m _{ \tilde{t} _1 } - \mchi < m _W + m _b \approx 85 \; \mbox{GeV} $. This condition defines what we call the `compressed' scenarios, in which the only two-body decay available to the heavy stop is into $\tone Z$. The $\ttwo$ BRs are therefore insensitive to $c_\thetat$ in this regime. This parameter does control the mass, and therefore also the decays of $\bone$. A heavy $\bone$ can decay into $\tone W^-$, but if this is not available it will have $\sim 100 \%$ BR into $b \neut$. For generic mass splittings this decay is highly constrained by dedicated searches, but having chosen small splittings for $\tone-\neut$ and $\ttwo-\tone$, this automatically places also the sbottom decay into the compressed regime in which this channel is particularly challenging for experimental searches. As we shall discuss in more detail in \Sec{sec:searches}, for a broad range of stop mixing angles the sbottom lies in a funnel of parameter space not constrained by these searches.

\subsection{Three Scenarios}

Four combinations of \fc versus \fv and split versus compressed are possible. Flavor conserving split (\fc-S) is the most highly constrained by dedicated searches, and we have been unable to find a region of its parameter space which permits an explanation of the $Z$ excess without being excluded by other searches. We therefore do not discuss this possibility in detail in this paper. Three combinations remain, which are summarized in \Tab{tab:scenarios}. For the compressed scenarios, we choose $\mttwo -\mtone  = 100 \; \text{GeV}$ and explore the $\mtone-\mchi$ plane. We set the branching ratios $\text{BR}(\ttwo \to \tone \neut) = 1$ and $\text{BR}(\bone \to b \neut) = 1$. In principle, $ \tilde{t} _2 $ also has competing three-body decays to $\tilde{b} _1 $, but these are sensitive to $ m _{ \tilde{b} _1} $ and would have a BR no more than $\mathcal{O}(5\%)$. We neglect this effect for simplicity. For the split scenario, we explore $\ttwo-\tone$ mass splittings between 125 and 300 GeV, choosing $c_\thetat = 0.5$ which is close to the optimal value for maximizing $\text{BR}(\ttwo \to \tone Z)$ over most of the parameter space. This angle then also sets the BRs for the $\bone$. We also compute the $\ttwo \to t \neut$ and $\bone \to b \neut$ BRs assuming a Bino LSP.

\begin{table}
\begin{tabular}{c | c c c}
Scenario & $\tone$ decay & $\mttwo - \mtone$ & BR($\ttwo \to \tone Z$) \\ \hline
\fcc & $f f' b \neut$ & 100 GeV & 1\\
\fvc & $c \neut$ / $u \neut$ & 100 GeV & 1\\
\fvs & $c \neut$ / $u \neut$ & (125 -- 300) GeV & $0.7 \pm 0.1$
\end{tabular}
\caption{The three scenarios considered in this paper, labeled Flavor Conserving Compressed, Flavor Violating Compressed, and Flavor Violating Split. The \fc/\fv designation refers to the decays of $\tone$, and the compressed/split designation refers to the splitting between $\ttwo$ and $\tone$.}
\label{tab:scenarios}
\end{table}

\section{Relevant searches}
\label{sec:searches}
ATLAS and CMS have a wealth of searches looking for large MET with all types of additional particles in the final state, each potentially providing a limit on stop and sbottom production. Since in general $ \tilde{t} _1 , \tilde{t} _2 , $ and $ \tilde{b} _1 $ will contribute to each bound, the constraints needs to be recast with care. Our goal is to examine all the parameter space with the simplified topology discussed in \Sec{sec:Model}. For the compressed cases this involves a scan in the $ \mchi - m _{ \tilde{t} _1 } $ plane while for the split case it involves a scan in both $ \mchi - m _{ \tilde{t} } $ and $ m _{ \tilde{t} _1 } - m _{ \tilde{t} _2 } $. We will be exploring scenarios with compressed mass splittings, especially between $\tone$ and $\neut$. This is a region that is very challenging experimentally. The most robust and model-independent limits in the most compressed regime come from dedicated searches for events with a hard jet coming from Initial State Radiation (ISR), which do not depend sensitively on the details of the $\tone$ decay. Searches for jets and $\ETmiss$ are highly constraining for spectra with a large splitting between the LSP and colored particles, but are challenging to interpret in the compressed regime where there are large systematic uncertainties. Searches involving $b$-tagged jets place important limits on some of our decay channels. Finally, there are dedicated searches for events containing $Z$ bosons which could be sensitive to our model. We discuss the details of these searches in the following subsections, beginning with the ATLAS on-$Z$ search with a $3\sigma$ excess. 

\subsection{ATLAS on-$Z$}\label{sec:ATLASonZ}
The ATLAS on-$ Z $ search looked for final states with two leptons with invariant mass around the $Z$-pole, $ \ETmiss > 225 \; \text{GeV} $, and $ H _T \equiv p _T ^{ \ell _1 } + p _T ^{ \ell _2 } + \sum _{ i \in \text{jets}  }p _T ^{ _i } > 600 \; \text{GeV} $. The $ H _T $ and $ \ETmiss $ cuts pick out events with large kinetic energies. ATLAS found $ 16 $ events ($ 4.2 \pm 1.6 $ expected) in the electron channel and $ 13 $ events ($ 6.4 \pm 2.2 $ expected) in the muon channel for a total of $ 29 $ events ($ 10.6 \pm 3.2 $ expected). Running pseudo-experiments they concluded that this corresponds to a $ 3.0 \sigma $ deviations from the SM. 

To estimate the number of events needed to explain the signal we use a log-likelihood method and profile over the background uncertainties using a Gaussian approximation. Using asyptotic formulae \cite{Cowan:2010js} to establish two-sided convidence limits (CL), we find that a minimum of 7.1 (12.4) signal events are required to be consistent with the excess at the 95\% (68\%) CL.

\subsection{CMS on-$Z$\label{sec:CMSonZ}}
CMS performed a search analogous to that done by ATLAS, looking for events with opposite-sign same-flavor leptons and $\ETmiss$~\cite{CMS_Z} which provides an important constraint on our model. CMS split their on-$ Z$ signal regions into six bins, depending on jet multiplicity and $\ETmiss$. Three bins measure events with $n_\text{jets} \ge 2$, and three use $n_\text{jets} \ge 3$. Our signal model does not produce a significant number of 2-jet events, and we therefore place constraints on our scenarios using the 3-jet inclusive bins which have higher expected sensitivity. The three bins in this category are split into the following $\ETmiss$ windows:  $ 100 \; \text{GeV} < \ETmiss < 200 \; \text{GeV}  $, $ 200 \; \text{GeV} < \ETmiss < 300 \; \text{GeV}  $, $ \ETmiss > 300 \; \text{GeV} $. We use the most constraining of these bins to constrain our scenarios at each point in parameter space, using the 95\% CL$_s$ limit to set bounds \cite{Read:2002hq}. In our scans (described in \Sec{sec:scan}) We found that the mid $\ETmiss$ bin usually provides the dominant limit, but the high $\ETmiss$ bin is sometimes competitive. The low $\ETmiss$ bin is never competitive with the other two in our simulations.\footnote{A possible concern is that the combined limit from different bins could be more severely constraining.  We find that usually only one bin is constraining and we do not expect a combination to notably alter the limits.}

\subsection{$ \tilde{t} _2 \rightarrow \tilde{t} _1 Z $}
Searches for the signal $ \tilde{t} _2 \rightarrow \tilde{t} _1 Z $ at $ 8~\text{TeV} $ have been performed both by ATLAS~\cite{ATLAS_tZt} and CMS~\cite{CMS_tZt}. The searches have competitive bounds, with ATLAS having a slightly better exclusion. For simplicity we only use the ATLAS search to place bounds on the \fcc scenario. The main cuts in this search are on the invariant mass of the leptons (which are required to be around the $ Z $ pole), the number of jets, number of leptons, and the $ p _T $ of the reconstructed $ Z $. We find that the reach of the search is limited in the compressed regime, both due to the small $\ttwo\text{--}\tone$ splitting and the small $\tone\text{--}\neut$ splitting. Firstly, the search regions requiring two leptons require a boosted $Z$ candidate which is suppressed in the \fcc scenario by the small $\ttwo\text{--}\tone$ splitting. Secondly, we find that the soft leptons and $b$-jets coming from the decay of $\tone$ in the 4-body regime reduce the acceptance in the 3-lepton bins and the $b$-tagging efficiency. Having recasted this search for the \fcc scenario, we find that it does not place competitive limits and we therefore omit it from our scans. We note that the CMS search is optimised for mass spitting $\mtone - \mchi \simeq m_t$ and has potential sensitivity to our \fcc scenario only in its 3-lepton bins, and will therefore also have degraded sensitivity in the 4-body region.

\subsection{Jets+MET+0/1 lepton}
The MSSM-inspired jets+MET searches\footnote{It was pointed out in~\cite{Agrawal:2013kha} that searches using shape-based analyses might have better sensitivity for $ \tilde{t} _1 \rightarrow j \neut $ then jets+MET in the limit of small $ \mtone - \mchi $. The bounds were computed using $ 7 \; \text{TeV} $ data and are not constraining compared to the $8 \; \text{TeV} $ jets+MET search. It would be interesting to see how these would change with the full $8 \; \text{TeV} $ data set.}
 provide an important constraint on our scenarios as $ \tilde{t} _1 , \tilde{t} _2 , $ and $ \tilde{b} _1 $ can all contribute to this signal. ATLAS~\cite{Aad:2014wea} and CMS~\cite{Chatrchyan:2014lfa} have both performed searches for this signature at 8 TeV and interpreted them in terms of a variety of SUSY models, including direct production of squarks decaying via $ \tilde{q} \rightarrow j \neut  $ (which is identical to $\tone$ in the \fv scenarios).  In this work we focus on the $ 200 \; \text{GeV} \lesssim m _{ \tilde{t} _1 } \lesssim  350 \; \text{GeV} $ region but CMS only presents limits for $ m _{ \tilde{q} } \gtrsim 300 \; \text{GeV} $. Thus to study the bounds from these searches we use the ATLAS analysis. Jets $ + $ MET searches in these regions are particularly challenging due to the low $ p _T $ of the outgoing particles. As a consequence their search does not constrain single squark production decaying in this channel for $\mchi > 160 \; \text {GeV}$. Another key factor which limits the reach of this search in the compressed regime is the systematic uncertainty on the ($\text{acceptance} \times \text{efficiency}$) associated with the high sensitivity to ISR. ATLAS provides uncertainties for each signal region in their auxiliary material (available on the ATLAS public results website), and these range from 10\% to 50\% in constraining bins.

In order to interpret the results of this search in terms of limits on our scenarios, it is necessary to combine the contributions from $\ttwo$, $\tone$ and $\bone$. Each of these channels will come with its own set of uncertainties which vary from bin-to-bin. Without a dedicated detector simulation it is not possible to robustly account for these effects. Instead, we estimate reasonable sizes for these uncertainties and study the effects of varying these assumptions. The ATLAS collaboration also interpreted their results in terms of some multi-step decay chains, and find uncertainties that range from 10\% to 80\% in similarly compressed regimes. Using the `$r$' method\footnote{In the $r$-method, a signal is excluded if $r \equiv (S - 1.96 \Delta S)/S_{95}^\text{obs} > 1$, where $\Delta S$ is the systematic uncertainty on the signal strength and $S_{95}^\text{obs}$ is the limit on the signal strength at the 95\% confidence limit.}~\cite{Drees:2013wra}, we find good agreement with the ATLAS exclusion on $\tone$ if we assign a uniform systematic uncertainty of 30\% on its signal strength across all signal regions  We then assign a nominal 30\% uncertainty also on the $\ttwo$ and $\bone$ contributions, consistent with the aforementioned uncertainties quoted by the ATLAS collaboration for other compressed multi-step decay processes. In order to asses the sensitivity of our results to this choice, we also vary the uncertainty on the $\ttwo$ and $\bone$ channels to 20\% and 40\%. We used CheckMATE~\cite{Drees:2013wra} to apply this technique to all our scenarios however we found the bounds of this search to be subdominant in all cases except for the \fvs case. This is because this is the only scenario in which we venture close to the existing bounds on  $ \tilde{q} \rightarrow j \neut  $, where a combination with the other channels might then result in an exclusion.

In addition to jets+MET there are searches which require an additional isolated lepton by both ATLAS~\cite{Aad:2014kra} and CMS~\cite{Chatrchyan:2013xna}, which are potentially sensitive to the \fcc scenario. However, the limits on the light stop are weaker than the other limits which we consider for this search, and we find that the heavy stop production does not contribute significantly to this signal. We therefore do not include this limit in our scans.

\subsection{Sbottom bounds}
The sbottom mass is determined by the mixing angle as given in \Equ{eq:mb1}. The phenomenology of the sbottom differs substantially between the compressed and split scenarios. In the compressed scenario, the only two-body final state available for it to decay into is $b \neut$. Furthermore, the mixing angle is a free parameter since the branching ratio of $ \tilde{t} _2 \rightarrow \tilde{t} _1 Z $ is fixed to $1$ by the kinematics. On the other hand, in the split scenario this mixing angle is fixed at $c_\thetat = 0.5$ and the sbottom decays almost exclusively into $\tone W$ in most of the parameter space.

The $\bone \to b \neut$ decay of the compressed scenarios is constrained by dedicated CMS and ATLAS searches~\cite{CMS:2014nia,Aad:2013ija}, and we focus on the ATLAS analysis because it places stronger limits in the compressed regime. This search places strong limits on this channel, but allows for a sbottom with mass $\mbone \simeq 250 \; \text{GeV}$ if it has a small mass splitting with the neutralino. We also require that the sbottom is heavy enough to forbid the decay $ \tilde{t} _2 \rightarrow \tilde{b} _1 W $, i.e., $ m _{ \tilde{t} _2 } < m _{ \tilde{b} _1 } + m _W $ or in terms of $ \tilde{t} _2 - \tilde{t} _1 $ mass splitting, $ m _{ \tilde{b} _1 } > m _{ \tilde{t} _1 } + \Delta m _{ \tilde{t} } - m _W $. Combining the bounds gives

\begin{equation} 
\mtone < m _{ _{\tilde b} } ^{ \text{lim}} ( \mchi ) - \Delta m_{\tilde{t}} + m _W 
\end{equation} 
where $ m _{ \tilde{b} } ^{ \lim} ( \mchi ) $ is the maximum allowed value of the sbottom mass for each $ \mchi $, coming from the ATLAS limit. This provides an additional constraint on the possible values in the $ \mchi , m _{ \tilde{t} _1 } $ plane. The additional $\tone$, $\ttwo$ production channels are not expected to contribute significantly to this search.

In the split scenario, \fvs, the sbottom decays predominantly in $ \tilde{b} _1 \rightarrow \tilde{t} _1 W \rightarrow j \neut W $. While there are no direct searches for this signal, there are searches for $ \tilde{q} \rightarrow \chi ^\pm W \rightarrow j \neut W $. We have checked the constraints due to this signal and found that we are well within experimental bounds for all regions of parameter space.

\subsection{Single high $ p _T $ jet + $ 0,1,2 $ lepton}
In this work we are primarily interested in regions of parameter space with small $\tilde{t} _1- \neut $ splitting where many jets may not pass the $ p _T $ cuts. In this case we have additional constraints from monojet searches. This search has been done by ATLAS for both $ \tilde{t} _1 \rightarrow j \neut $ and $ \tilde{t} _1 \rightarrow b f f W $ stop decay modes in a search by ATLAS~\cite{Aad:2014nra}. For the split case $ \tilde{t} _1 $ is the only production channel which can replicate this signal, but in the compressed cases the $ \tilde{b} _1  \rightarrow b \neut $ could provide additional monojet events. To this end we recast the search including just $ \tilde{t} _1 $ production and both $ \tilde{t} _1 $ and $ \tilde{b} _1 $ but we found comparable exclusions. For this reason we simply include the constraints computed by ATLAS directly in our analysis. 

CMS has performed a search for events with a high momentum ISR jet with the additional requirement of one or two soft leptons \cite{CMS:2015eoa}. The preliminary results of this search provide the strongest existing constraints on the 4-body region of flavor-conserving $\tone$ decays, sensitive to stop masses up to $320 \; \text{GeV}$. The strongest bounds are derived from the 2-lepton signal region, and we therefore expect this limit to be highly sensitive to the $\tone$ BR.

\subsection{Charm-tagging}
A final constraint on our signals are charm-tagging searches. ATLAS has two searches that employ charm-tagging looking for both $ \tilde{t} _1 \rightarrow c \neut$~\cite{Aad:2014nra} as well as $ \tilde{c} \rightarrow c \neut $~\cite{Aad:2015gna}. The $ \tilde{t} _1 \rightarrow c \neut $ search assumes the stop is in the four-body regime and hence is optimized for our signal. For this reason it has better sensitivity in our region of interest. For this reason we omit it from our plots. 

The c-tagging searches put constraints on \fv models which involve charm. For the \fvs split scenatrio we find that these constraints rule out the region preferred by the ATLAS on-$Z$ excess at the 95\% CL. For this reason in this scenario we assume $ \tilde{t} _1 $ decays to $ u \neut $. For the \fvc scenario the constraints are milder since we are exploring relatively large $ m _{ \tilde{t} _1 } $ values. Recasting the $ \tilde{t} _1 \rightarrow c \neut $ search we find that the limits on this scenario comparable to those directly on $ \tilde{t}  _1 $ alone, and thus we use the limits on this channel that are provided in~\cite{Aad:2014nra}.

\begin{figure*}
\includegraphics[width=0.85\linewidth]{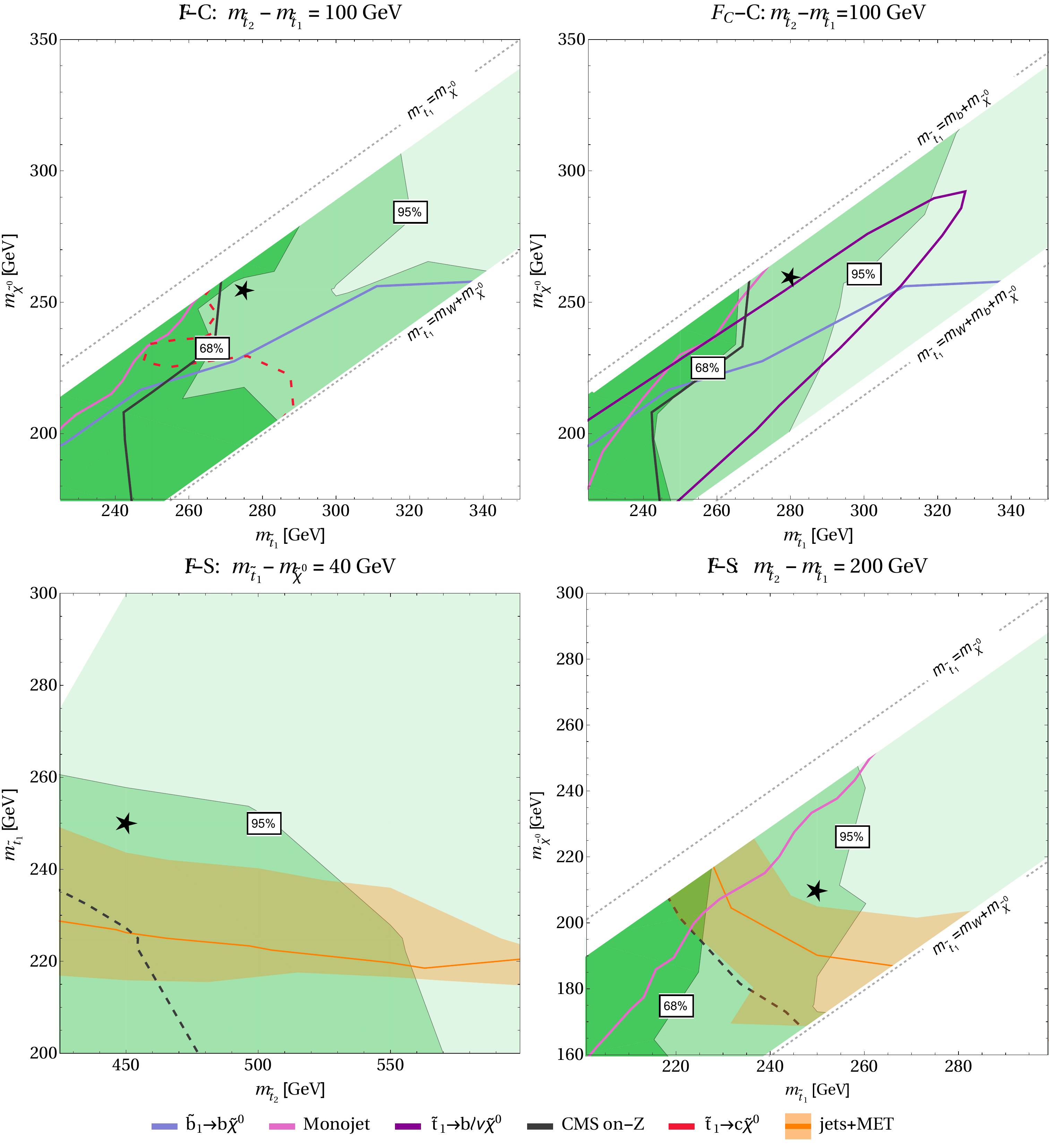}
\caption{Scans of the three scenarios. The regions favoured by the ATLAS on-$Z$ excess are shaded green, with contours indicating 95\% and 68\% confidence intervals. Additional solid lines indicate the 95\% limits described in the text. The dashed lines indicate limits under specific model assumptions that do not necessarily apply, as described in the text. The band on the jets+MET limit illustrates the considerable uncertainty on the strength of this limit. Black stars indicate the benchmark points chosen from each scenario, and they also indicate the region of parameter space that is not excluded by the other searches.}
\label{fig:scan}
\end{figure*} 

\section{Scan}
\label{sec:scan}
For the scan we use Madgraph 5 v2.2.3~\cite{Alwall:2011uj}, Pythia 6.4~\cite{Pythia}, and PGS~\cite{PGS}, including 1-jet matching. For jet clustering we use anti-kT algorithm with $ \Delta R = 0.4 $. To roughly account for next-to-leading order (NLO) effects we rescale our cross sections to their NLO values calculated by the SUSY Cross Sections group~\cite{LHCcrosssections}. For jets+MET and double-checking monojet constraints we use CheckMATE~\cite{Drees:2013wra}, which makes use of DELPHES 3 \cite{deFavereau:2013fsa}, FastJet~\cite{Cacciari:2011ma}, and the anti-kT clustering algorithm~\cite{Cacciari:2008gp}. For the compressed scenarios, \fcc and \fvc, there is one less free parameter (since the range of $ m _{ \tilde{t} _2 } $ has a relatively small range of viable options). We can perform a two dimensional scan over $ m _{ \tilde{t} _1 } - \mchi $. \fvs requires a $3$ dimensional scan but for simplicity we scan over two slices in the parameter space.

The scan showing the signal as well as limits from the different searches is shown in \Fig{fig:scan}, with the regions preferred by the ATLAS on-$Z$ excess indicated by green shading and the contours labelling the 90\% and 68\% two-sided confidence intervals. The constraining 95\% confidence intervals discussed in \Sec{sec:searches} are shown by solid lines. The dashed line in the \fvc scan indicates the limit on the decay $\tone \to c \neut$, though the alternate decay $\tone \to u \neut$ is also possible and not constrained by this line. The CMS on-$Z$ limit in the \fvs scenarios is dashed as it has been calculated assuming no background contamination. As we shall discuss below, considerable background contamination is expected which severely limits the sensitivity of the CMS search to this scenario. The jets+MET limit in the \fvs scenario is plotted with a band indicating the large uncertainties associated with this search in the compressed regime, as discussed in \Sec{sec:Model}. The central line assumes a systematic uncertainty on the signal from all channels and in all bins of 30\%. The band is obtained by varying the uncertainty on the $\tone$ and $\bone$ production channels to 20\% and 40\%.

These plots indicate that all three scenarios can be consistent with the ATLAS on-$Z$ excess at the 90\% level and the two compressed scenarios at the $1\sigma$ level, allowing for as many as 14 signal events. From each scenario we have chosen a benchmark point indicated by a black star in \Fig{fig:scan}, and detailed in \Tab{tab:benchmarks}.

\begin{table}
\begin{tabular}{c||c|c|c|c}
Benchmark & $\mttwo$ [GeV] & $\mtone$ [GeV] & $\mchi$ [GeV]  & p-value\\ \hline
\fcc & 380 & 280 & 260 & 0.095\\
\fvc & 370 & 275 & 255 & 0.17\\
\fvs & 450 & 250 & 210  & 0.055
\end{tabular}
\caption{Benchmark points chosen from the three scenarios. All other parameters are as described in \Sec{sec:Model}. The two-tailed $p$-values are calculated as described in \Sec{sec:ATLASonZ}, and a $p$-value of 1  would represent perfect agreement with the measured total event rate.}
\label{tab:benchmarks}
\end{table}

\subsection{Kinematic distributions}
\begin{figure*} 
\includegraphics[width=0.95\linewidth]{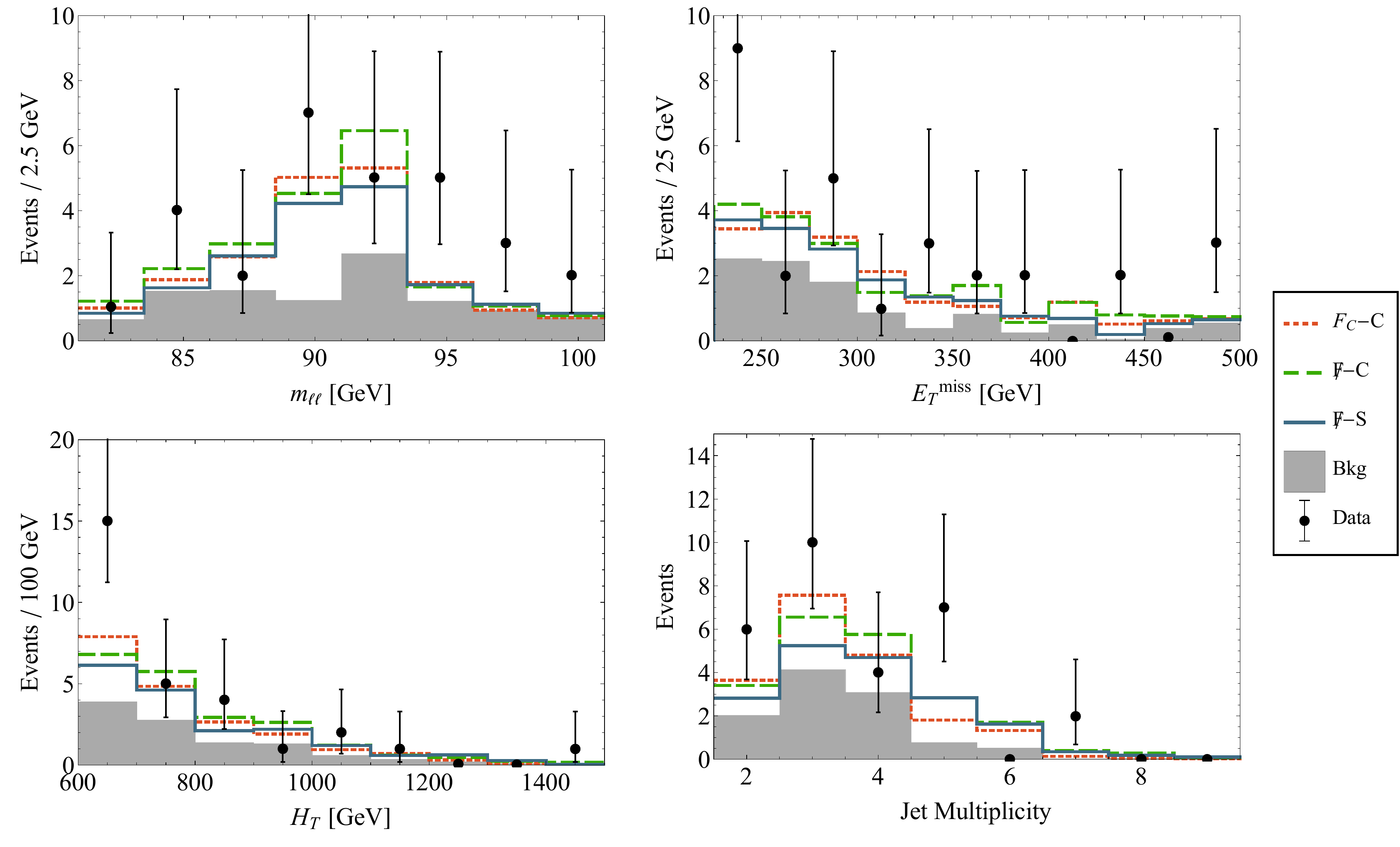} 
\caption{The kinematic distributions compared to those of ATLAS in the on-$Z$ search. The simulation predictions are summed with the ATLAS-calculated SM background to produce the model predictions.}
\label{fig:fig67}
\end{figure*} 
To further check the consistency with the data we compare our $ m _{ \ell \ell} ,\ETmiss , H _T , $ and $ n _{\text{jets}} $ distributions with those measured by the ATLAS on-$Z$ search (these correspond to Fig.6 and 7 in~\cite{ATLAS_Z}). To compare the quality of our signal we reproduce these plots in \Fig{fig:fig67} using the results by ATLAS to retrieve the SM background.
We see good agreement across all kinematic variables. In particular, unlike for other viable models which tend to peak at high number of jets (large numbers of jets is often accompanied by large $ H _T $), we can roughly reproduce the jet multiplicity plot distributions. If the excess persists this could be a powerful variable to discriminate between candidate interpretations. We also note that this signal peaks at values for $H _T$ and $E _T ^{ \text{miss}} $ far below the the thresholds for the kinematic cuts of the ATLAS search, as opposed to models based on the cascade decays of much heavier particles that have been previously considered. Nonetheless, we still evade the bounds from the CMS search which has weaker $ E _T ^{ \text{miss}} $ cuts, due to the sharp increase in the background. For instance, the CMS background estimates were $ 478 \pm 43 $, $ 39.2 \pm 6.6 $, and $ 5.3 \pm 2.3 $ events in the $ 100 \text{ GeV} <E _T ^{\text{miss}} < 200 \text{ GeV}, 200 \text{ GeV} < E _T ^{\text{miss}} < 300 \text{ GeV}, $ and $ E _T ^{ \text{miss}} > 300 \text{ GeV} $ bins with $ n _{ jets} \ge 3 $, while for the ${\slashed F} - C $ benchmark we predict $ 35 $, $ 11 $, and $ 5.4 $ events respectively. We see that the large event rates in the low $ E _T ^{ \text{miss}} $ are well within the background uncertainties.

\subsection{Background contamination}
Another interesting feature of our signal is that it allows for the possibility of significant background contamination in the CMS search for the same final state described in \Sec{sec:CMSonZ}. One of the most significant backgrounds in this search comes from SM Drell Yan (DY) production of $Z$ bosons. To estimate this background, the CMS collaboration used two independent data-driven methods and took a weighted average. One of these methods is based on the variable `jet-$Z$ balance' (JZB) \cite{Buchmann:2012np, CMS:2011mya}, which is important particulatly in the high $\ETmiss$ search regions which constrain our signal. The JZB of an event is defined by
\begin{equation} 
\text{JZB} \equiv \big| \sum _{i\in \text{jets}} \vec{ p } _T ^{ i}  \big|  - \big| \vec{ p } _T ^{(Z)} \big| = \big| \vec{ E } _{ T} ^{\text{miss}}+ \vec{ p } _T ^{ Z} \big| - \big| \vec{ p } _T ^{ (Z) } \big|.
\end{equation} 
SM processes like DY production typically result in JZB distributions that are symmetric about $\text{JZB}=0 \; \text{GeV}$ (because a non-zero value arises from jet energy resolution effects), while some BSM processes can have JZB distributions that are strongly skewed towards positive values. This is expected when the $Z$ is emitted back-to-back with an invisible particle, e.g. in a decay chain ending in $ \neut _2  \rightarrow \neut _1  Z $. For this reason, the JZB method estimates the DY background by assuming all events with $\text{JZB} < 0 \; \text{GeV}$ are produced by DY, and extrapolating this to positive JZB values under the assumption that DY production is JZB-symmetric.

\begin{figure}
  \begin{center} 
\includegraphics[width=0.9\linewidth]{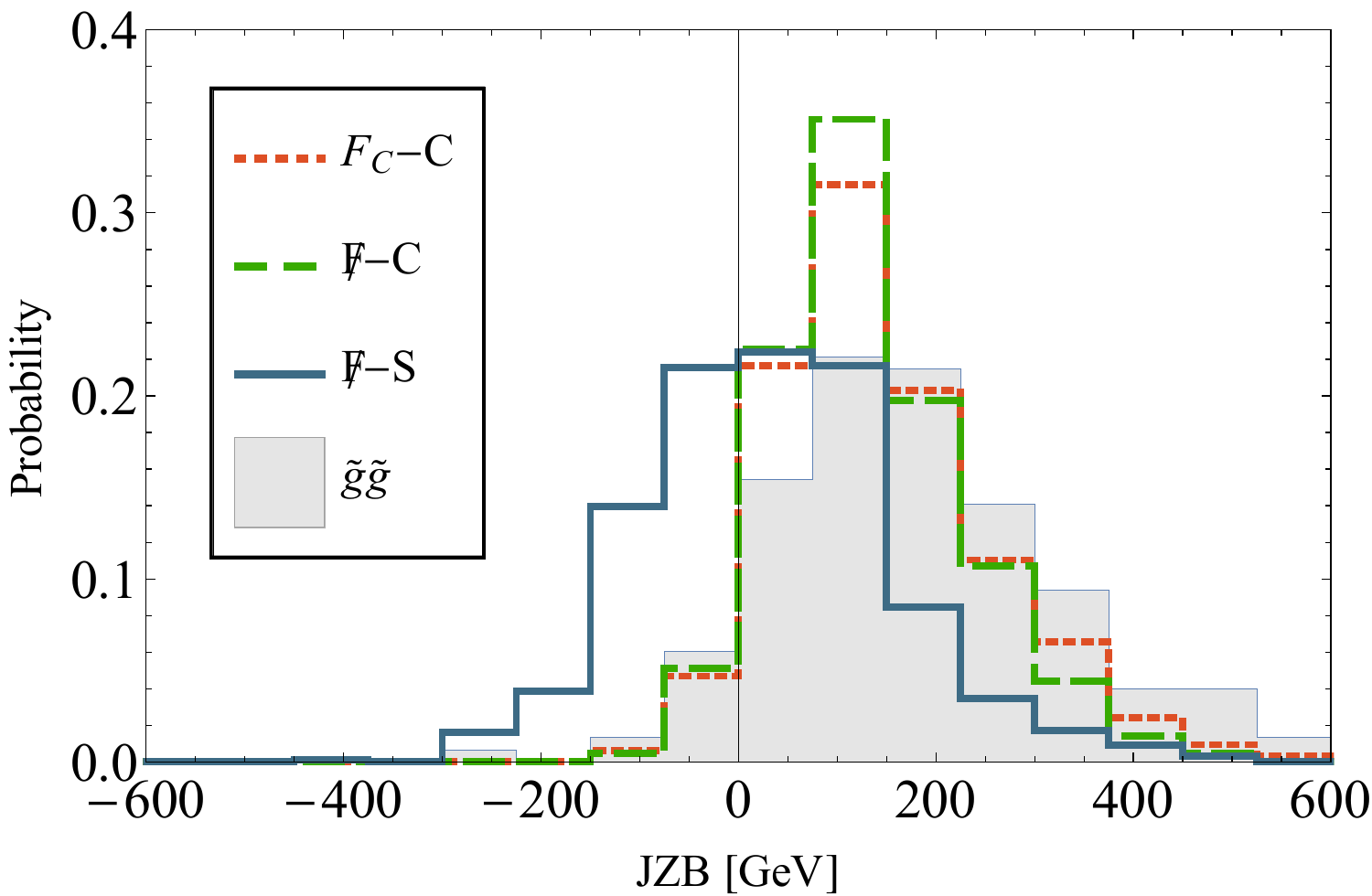} 
\caption{JZB probability distributions for the benchmark points when compared with a typical gluino production scenario ($\tilde{g}\to q q \chi^0_2 \to q q \chi^0_1 Z$). The gluino scenario has the most positively skewed JZB distribution, while the \fvs has almost symmetrical JZB.}
\label{fig:JZB}
\end{center}
\end{figure}

It is clear therefore that signals with symmetric JZB distributions would contaminate this background estimate, reducing the sensitivity of the CMS search. In \Fig{fig:JZB} we plot the JZB distributions after applying the cuts for the CMS $ n  _{ \text{jets}} \ge 3 $, mid $\ETmiss$ bin for our three benchmark points, as well as for a gluino production model with the decay $\tilde{g}\to q q \chi^0_2 \to q q \chi^0_1 Z$ (we have chosen the parameters $ m _{ \tilde{g} } = 950 \; \text{GeV}, m _{ \chi ^0 } = 50 \; \text{GeV}$). We find that the JZB distribution is highly sensitive to the $\tone \text{--} \ttwo$ mass splitting in this model. For small splitting, the $Z$ tends to be very soft resulting in positive JZB. For large splitting, the hard $Z$ can result in symmetric or even negatively skewed JZB distributions. The \fvs benchmark point has 41\% of events with $\text{JZB} < 0 \; \text{GeV}$, comparing with only 8\% in the gluino model. This also highlights the potential for the JZB distribution to be used as a discriminating variable between new physics explanations of this excess should it persist in the next run of the LHC, due to its sensitivity to where the $Z$ is emitted.

\section{Conclusion}
Motivated by the recent $3\sigma$ excess reported by the ATLAS collaboration in a $Z+\text{jets}+E_T^\text{miss}$ channel we have studied if it can potentially be explained in the context of a natural supersymmetric spectrum involving light stops. Strong constraints on such scenarios have led us to a compressed spectrum featuring two light and mixed stops and a light LSP. We identified three possible scenarios, characterized by flavor conserving or flavor violating decays of the lightest stop, and the splitting between the two stop masses. We have shown that in all three scenarios it is possible to produce the excess within $2\sigma$, while in \fvc and \fcc we can reproduce the excess within $1 \sigma$ of the ATLAS measurement. While the scenarios should be taken as examples, it is clear that possible interpolations between them are capable of addressing the excess and would retain the same general features. Such features are a light stop with $225 \; \text{GeV} \lesssim \mtone \lesssim 325 \; \text{GeV}$, almost degenerate with a Bino-like LSP and mixed with a second light stop with mass $325 \; \text{GeV} \lesssim \mttwo \lesssim 550 \; \text{GeV}$. 

The topology of the process differs from previous attempts to address the excess. The most substantial difference is the production of the $Z$'s in the first step of a decay chain, and not in the last step in association with an invisible particle responsible for the $E_T^\text{miss}$. Interestingly, we have shown that it could lead to the contamination of background estimation based on the JZB method. This method is employed in a CMS search for a similar final state. We have estimated that as much as half of the signal could fall in the background control region, which could lead to over-exclusions. Additionally, we notice that the JZB variable could be used to discriminate between different signal topologies if this excess turns out to be due to new physics.

Should this excess persist in run-II, it will be crucial to distinguish between the signal hypotheses. The signature proposed in this work is distinguished by its light compressed spectrum. This resulted in monojet searches being a highly sensitive probe of our signal. In addition, the search for $ \tilde{b} _1  \rightarrow b \neut $ is highly complementary, and between these searches the region of parameter space which can explain the excess should be fully explored at $ 13 \text{ TeV} $.

We note in passing that there are additional modest excesses of around two sigma or more in final states containing b-jets, leptons and MET, including a 1.9 sigma `on-$Z$' excess in events with low jet multiplicity~\cite{CMS:2013ida}, and various hints of same-sign dileptons with b-jets and MET (see~\cite{Huang:2015fba} for a summary). Light stops and sbottoms can give rise to all of these signatures, and it is interesting to consider the possibility that if these really are all hints of new physics, they could have a unified explanation in a more complete model. Whether the ATLAS excess is a fluctuation or a first tantalizing hint of new physics will soon be decided.

\section*{Acknowledgments}
We are thankful to Ennio Salvioni, Jong Soo Kim, Jamie Tattersall, Tommaso Lari, Ben Allanach, Csaba Cs\'aki and Maxim Perelstein for useful discussions. The work of JAD is supported in part by NSERC Grant PGSD3-438393-2013. The work of JHC is supported by the Cornell Graduate Student Fellowship. This research is supported by the U.S. National Science Foundation through grant PHY-1316222.

\bibliographystyle{JHEP}
\bibliography{bibZexcess.v6}{}

\providecommand{\href}[2]{#2}\begingroup\raggedright\begin{thebibliography}{10}

\bibitem{ATLAS_Z}
{\bf ATLAS} Collaboration, G.~Aad et~al., {\it {Search for supersymmetry in
  events containing a same-flavour opposite-sign dilepton pair, jets, and large
  missing transverse momentum in $\sqrt{s}=8$ TeV pp collisions with the ATLAS
  detector}},  {\em Eur. Phys. J.} {\bf C75} (2015), no.~7 318,
  [\href{http://arxiv.org/abs/1503.03290}{{\tt arXiv:1503.03290}}].

\bibitem{Ghosh:2013qga}
D.~Ghosh, {\it {Boosted dibosons from mixed heavy top squarks}},  {\em Phys.
  Rev.} {\bf D88} (2013), no.~11 115013,
  [\href{http://arxiv.org/abs/1308.0320}{{\tt arXiv:1308.0320}}].

\bibitem{Vignaroli:2015ama}
N.~Vignaroli, {\it {$Z$-peaked excess from heavy gluon decays to vectorlike
  quarks}},  {\em Phys. Rev.} {\bf D91} (2015), no.~11 115009,
  [\href{http://arxiv.org/abs/1504.01768}{{\tt arXiv:1504.01768}}].

\bibitem{Ellwanger:2015hva}
U.~Ellwanger, {\it {Possible explanation of excess events in the search for
  jets, missing transverse momentum and a Z boson in pp collisions}},
  \href{http://arxiv.org/abs/1504.02244}{{\tt arXiv:1504.02244}}.

\bibitem{Cao:2015ara}
J.~Cao, L.~Shang, J.~M. Yang, and Y.~Zhang, {\it {Explanation of the ATLAS
  Z-Peaked Excess in the NMSSM}},  {\em JHEP} {\bf 06} (2015) 152,
  [\href{http://arxiv.org/abs/1504.07869}{{\tt arXiv:1504.07869}}].

\bibitem{Allanach:2015xga}
B.~Allanach, A.~Raklev, and A.~Kvellestad, {\it {Consistency of the recent
  ATLAS $Z+E\_T^{\rm miss}$ excess in a simplified GGM model}},  {\em Phys.
  Rev.} {\bf D91} (2015) 095016, [\href{http://arxiv.org/abs/1504.02752}{{\tt
  arXiv:1504.02752}}].

\bibitem{Barenboim:2015afa}
G.~Barenboim, J.~Bernabeu, V.~A. Mitsou, E.~Romero, E.~Torro, and O.~Vives,
  {\it {METing SUSY on the Z peak}},
  \href{http://arxiv.org/abs/1503.04184}{{\tt arXiv:1503.04184}}.

\bibitem{Kobakhidze:2015dra}
A.~Kobakhidze, A.~Saavedra, L.~Wu, and J.~M. Yang, {\it {ATLAS Z-peaked excess
  in MSSM with a light sbottom or stop}},
  \href{http://arxiv.org/abs/1504.04390}{{\tt arXiv:1504.04390}}.

\bibitem{Cahill-Rowley:2015cha}
M.~Cahill-Rowley, J.~L. Hewett, A.~Ismail, and T.~G. Rizzo, {\it {The ATLAS Z +
  MET Excess in the MSSM}},  \href{http://arxiv.org/abs/1506.05799}{{\tt
  arXiv:1506.05799}}.

\bibitem{Lu:2015wwa}
X.~Lu, S.~Shirai, and T.~Terada, {\it {ATLAS Z Excess in Minimal Supersymmetric
  Standard Model}},  \href{http://arxiv.org/abs/1506.07161}{{\tt
  arXiv:1506.07161}}.

\bibitem{Liew:2015hsa}
S.~P. Liew, A.~Mariotti, K.~Mawatari, K.~Sakurai, and M.~Vereecken, {\it
  {Z-peaked excess in goldstini scenarios}},
  \href{http://arxiv.org/abs/1506.08803}{{\tt arXiv:1506.08803}}.

\bibitem{Cao:2015zya}
J.~Cao, L.~Shang, J.~M. Yang, and Y.~Zhang, {\it {Explanation of the ATLAS
  Z-peaked excess by squark pair production in the NMSSM}},
  \href{http://arxiv.org/abs/1507.08471}{{\tt arXiv:1507.08471}}.

\bibitem{CMS_Z}
{\bf CMS} Collaboration, V.~Khachatryan et~al., {\it {Search for physics beyond
  the standard model in events with two leptons, jets, and missing transverse
  momentum in pp collisions at sqrt(s) = 8 TeV}},  {\em JHEP} {\bf 04} (2015)
  124, [\href{http://arxiv.org/abs/1502.06031}{{\tt arXiv:1502.06031}}].

\bibitem{Blanke:2013uia}
M.~Blanke, G.~F. Giudice, P.~Paradisi, G.~Perez, and J.~Zupan, {\it {Flavoured
  Naturalness}},  {\em JHEP} {\bf 1306} (2013) 022,
  [\href{http://arxiv.org/abs/1302.7232}{{\tt arXiv:1302.7232}}].

\bibitem{Backovic:2015rwa}
M.~Backović, A.~Mariotti, and M.~Spannowsky, {\it {Signs of Tops from Highly
  Mixed Stops}},  {\em JHEP} {\bf 06} (2015) 122,
  [\href{http://arxiv.org/abs/1504.00927}{{\tt arXiv:1504.00927}}].

\bibitem{Agrawal:2013kha}
P.~Agrawal and C.~Frugiuele, {\it {Mixing stops at the LHC}},  {\em JHEP} {\bf
  01} (2014) 115, [\href{http://arxiv.org/abs/1304.3068}{{\tt
  arXiv:1304.3068}}].

\bibitem{Boehm:1999bj}
C.~Boehm, A.~Djouadi, and M.~Drees, {\it {Light scalar top quarks and
  supersymmetric dark matter}},  {\em Phys. Rev.} {\bf D62} (2000) 035012,
  [\href{http://arxiv.org/abs/hep-ph/9911496}{{\tt hep-ph/9911496}}].

\bibitem{Balazs:2004bu}
C.~Balazs, M.~Carena, and C.~E.~M. Wagner, {\it {Dark matter, light stops and
  electroweak baryogenesis}},  {\em Phys. Rev.} {\bf D70} (2004) 015007,
  [\href{http://arxiv.org/abs/hep-ph/0403224}{{\tt hep-ph/0403224}}].

\bibitem{Martin}
S.~P. Martin, {\it {A Supersymmetry primer}},
  \href{http://arxiv.org/abs/hep-ph/9709356}{{\tt hep-ph/9709356}}. [Adv. Ser.
  Direct. High Energy Phys.18,1(1998)].

\bibitem{Aad:2015pfx}
{\bf ATLAS} Collaboration, G.~Aad et~al., {\it {ATLAS Run 1 searches for direct
  pair production of third-generation squarks at the Large Hadron Collider}},
  \href{http://arxiv.org/abs/1506.08616}{{\tt arXiv:1506.08616}}.

\bibitem{D'Ambrosio:2002ex}
G.~D'Ambrosio, G.~F. Giudice, G.~Isidori, and A.~Strumia, {\it {Minimal flavor
  violation: An Effective field theory approach}},  {\em Nucl. Phys.} {\bf
  B645} (2002) 155--187, [\href{http://arxiv.org/abs/hep-ph/0207036}{{\tt
  hep-ph/0207036}}].

\bibitem{Muhlleitner:2011ww}
M.~Muhlleitner and E.~Popenda, {\it {Light Stop Decay in the MSSM with Minimal
  Flavour Violation}},  {\em JHEP} {\bf 04} (2011) 095,
  [\href{http://arxiv.org/abs/1102.5712}{{\tt arXiv:1102.5712}}].

\bibitem{Grober:2014aha}
R.~Grober, M.~Muhlleitner, E.~Popenda, and A.~Wlotzka, {\it {Light Stop Decays:
  Implications for LHC Searches}},  \href{http://arxiv.org/abs/1408.4662}{{\tt
  arXiv:1408.4662}}.

\bibitem{Dedes:2015twa}
A.~Dedes, M.~Paraskevas, J.~Rosiek, K.~Suxho, and K.~Tamvakis, {\it {Mass
  Insertions vs. Mass Eigenstates calculations in Flavour Physics}},  {\em
  JHEP} {\bf 06} (2015) 151, [\href{http://arxiv.org/abs/1504.00960}{{\tt
  arXiv:1504.00960}}].

\bibitem{Cowan:2010js}
G.~Cowan, K.~Cranmer, E.~Gross, and O.~Vitells, {\it {Asymptotic formulae for
  likelihood-based tests of new physics}},  {\em Eur. Phys. J.} {\bf C71}
  (2011) 1554, [\href{http://arxiv.org/abs/1007.1727}{{\tt arXiv:1007.1727}}].
  [Erratum: Eur. Phys. J.C73,2501(2013)].

\bibitem{Read:2002hq}
A.~L. Read, {\it {Presentation of search results: The CL(s) technique}},  {\em
  J. Phys.} {\bf G28} (2002) 2693--2704. [,11(2002)].

\bibitem{ATLAS_tZt}
{\bf ATLAS} Collaboration, G.~Aad et~al., {\it {Search for direct top squark
  pair production in events with a Z boson, b-jets and missing transverse
  momentum in sqrt(s)=8 TeV pp collisions with the ATLAS detector}},  {\em Eur.
  Phys. J.} {\bf C74} (2014), no.~6 2883,
  [\href{http://arxiv.org/abs/1403.5222}{{\tt arXiv:1403.5222}}].

\bibitem{CMS_tZt}
{\bf CMS} Collaboration, V.~Khachatryan et~al., {\it {Search for top-squark
  pairs decaying into Higgs or Z bosons in pp collisions at $\sqrt{s}$=8 TeV}},
   {\em Phys. Lett.} {\bf B736} (2014) 371--397,
  [\href{http://arxiv.org/abs/1405.3886}{{\tt arXiv:1405.3886}}].

\bibitem{Aad:2014wea}
{\bf ATLAS} Collaboration, G.~Aad et~al., {\it {Search for squarks and gluinos
  with the ATLAS detector in final states with jets and missing transverse
  momentum using $\sqrt{s}=8$ TeV proton--proton collision data}},  {\em JHEP}
  {\bf 09} (2014) 176, [\href{http://arxiv.org/abs/1405.7875}{{\tt
  arXiv:1405.7875}}].

\bibitem{Chatrchyan:2014lfa}
{\bf CMS} Collaboration, S.~Chatrchyan et~al., {\it {Search for new physics in
  the multijet and missing transverse momentum final state in proton-proton
  collisions at $\sqrt{s}$= 8 TeV}},  {\em JHEP} {\bf 06} (2014) 055,
  [\href{http://arxiv.org/abs/1402.4770}{{\tt arXiv:1402.4770}}].

\bibitem{Drees:2013wra}
M.~Drees, H.~Dreiner, D.~Schmeier, J.~Tattersall, and J.~S. Kim, {\it
  {CheckMATE: Confronting your Favourite New Physics Model with LHC Data}},
  {\em Comput. Phys. Commun.} {\bf 187} (2014) 227--265,
  [\href{http://arxiv.org/abs/1312.2591}{{\tt arXiv:1312.2591}}].

\bibitem{Aad:2014kra}
{\bf ATLAS} Collaboration, G.~Aad et~al., {\it {Search for top squark pair
  production in final states with one isolated lepton, jets, and missing
  transverse momentum in $\sqrt s =$8 TeV $pp$ collisions with the ATLAS
  detector}},  {\em JHEP} {\bf 11} (2014) 118,
  [\href{http://arxiv.org/abs/1407.0583}{{\tt arXiv:1407.0583}}].

\bibitem{Chatrchyan:2013xna}
{\bf CMS} Collaboration, S.~Chatrchyan et~al., {\it {Search for top-squark pair
  production in the single-lepton final state in pp collisions at $\sqrt{s}$ =
  8 TeV}},  {\em Eur. Phys. J.} {\bf C73} (2013), no.~12 2677,
  [\href{http://arxiv.org/abs/1308.1586}{{\tt arXiv:1308.1586}}].

\bibitem{CMS:2014nia}
{\bf CMS} Collaboration, C.~Collaboration, {\it {Search for direct production
  of bottom squark pairs}}, .

\bibitem{Aad:2013ija}
{\bf ATLAS} Collaboration, G.~Aad et~al., {\it {Search for direct
  third-generation squark pair production in final states with missing
  transverse momentum and two $b$-jets in $\sqrt{s} =$ 8 TeV $pp$ collisions
  with the ATLAS detector}},  {\em JHEP} {\bf 1310} (2013) 189,
  [\href{http://arxiv.org/abs/1308.2631}{{\tt arXiv:1308.2631}}].

\bibitem{Aad:2014nra}
{\bf ATLAS} Collaboration, G.~Aad et~al., {\it {Search for pair-produced
  third-generation squarks decaying via charm quarks or in compressed
  supersymmetric scenarios in $pp$ collisions at $\sqrt{s}=8~$TeV with the
  ATLAS detector}},  {\em Phys. Rev.} {\bf D90} (2014), no.~5 052008,
  [\href{http://arxiv.org/abs/1407.0608}{{\tt arXiv:1407.0608}}].

\bibitem{CMS:2015eoa}
{\bf CMS} Collaboration, C.~Collaboration, {\it {Search for supersymmetry in
  events with soft leptons, low jet multiplicity, and missing transverse
  momentum in proton-proton collisions at sqrt(s) = 8 TeV}}, .

\bibitem{Aad:2015gna}
{\bf ATLAS} Collaboration, G.~Aad et~al., {\it {Search for Scalar Charm Quark
  Pair Production in $pp$ Collisions at $\sqrt{s}=$ 8  TeV with the ATLAS
  Detector}},  {\em Phys. Rev. Lett.} {\bf 114} (2015), no.~16 161801,
  [\href{http://arxiv.org/abs/1501.01325}{{\tt arXiv:1501.01325}}].

\bibitem{Alwall:2011uj}
J.~Alwall, M.~Herquet, F.~Maltoni, O.~Mattelaer, and T.~Stelzer, {\it {MadGraph
  5 : Going Beyond}},  {\em JHEP} {\bf 06} (2011) 128,
  [\href{http://arxiv.org/abs/1106.0522}{{\tt arXiv:1106.0522}}].

\bibitem{Pythia}
T.~Sjostrand, S.~Mrenna, and P.~Z. Skands, {\it {PYTHIA 6.4 Physics and
  Manual}},  {\em JHEP} {\bf 05} (2006) 026,
  [\href{http://arxiv.org/abs/hep-ph/0603175}{{\tt hep-ph/0603175}}].

\bibitem{PGS}
J.~Conway, {\it Pretty good simulation}, .

\bibitem{LHCcrosssections}
S.~Padhi et~al., {\it {LHC SUSY} cross sections working group},  2015.

\bibitem{deFavereau:2013fsa}
{\bf DELPHES 3} Collaboration, J.~de~Favereau et~al., {\it {DELPHES 3, A
  modular framework for fast simulation of a generic collider experiment}},
  {\em JHEP} {\bf 1402} (2014) 057, [\href{http://arxiv.org/abs/1307.6346}{{\tt
  arXiv:1307.6346}}].

\bibitem{Cacciari:2011ma}
M.~Cacciari, G.~P. Salam, and G.~Soyez, {\it {FastJet User Manual}},  {\em Eur.
  Phys. J.} {\bf C72} (2012) 1896, [\href{http://arxiv.org/abs/1111.6097}{{\tt
  arXiv:1111.6097}}].

\bibitem{Cacciari:2008gp}
M.~Cacciari, G.~P. Salam, and G.~Soyez, {\it {The Anti-k(t) jet clustering
  algorithm}},  {\em JHEP} {\bf 04} (2008) 063,
  [\href{http://arxiv.org/abs/0802.1189}{{\tt arXiv:0802.1189}}].

\bibitem{Buchmann:2012np}
{\bf CMS} Collaboration, M.-A. Buchmann, {\it {Search for supersymmetry in
  events with a Z boson, jets and missing energy}},  {\em EPJ Web Conf.} {\bf
  28} (2012) 12017, [\href{http://arxiv.org/abs/1201.3748}{{\tt
  arXiv:1201.3748}}].

\bibitem{CMS:2011mya}
{\bf CMS} Collaboration, C.~Collaboration, {\it {Search for Physics Beyond the
  Standard Model in Z + Jets + MET events at the LHC}}, .

\bibitem{CMS:2013ida}
{\bf CMS} Collaboration, C.~Collaboration, {\it {Search for supersymmetry in pp
  collisions at $\sqrt(s) = 8$ {TeV} in events with three leptons and at least
  one b-tagged jet}}, .

\bibitem{Huang:2015fba}
P.~Huang, A.~Ismail, I.~Low, and C.~E.~M. Wagner, {\it {Same-Sign Dilepton
  Excesses and Light Top Squarks}},
  \href{http://arxiv.org/abs/1507.01601}{{\tt arXiv:1507.01601}}.

\end{thebibliography}\endgroup

\end{document}